\documentclass[11pt]{spieman}  
\usepackage{amsmath,amsfonts,amssymb}
\usepackage{graphicx}
\usepackage{setspace}
\usepackage{tocloft}
\usepackage{txfonts}
\usepackage{subcaption} 

\usepackage{gensymb}

\makeatletter

\newcommand*{\rom}[1]{\expandafter\@slowromancap\romannumeral #1@}
\title{Polarization properties of X-ray tubes used for Imaging X-ray Polarimetry Explorer calibration}

\author[a,*]{Ajay Ratheesh}
\author[a]{John Rankin}
\author[a]{Enrico Costa}
\author[a]{Ettore Del Monte}
\author[a]{Alessandro Di Marco}
\author[a]{Sergio Fabiani}
\author[a]{Fabio La Monaca}
\author[a]{Fabio Muleri}
\author[a]{Alda Rubini}
\author[a]{Paolo Soffitta}
\author[b,c]{Luca Baldini}
\author[c]{Massimo Minuti}
\author[c]{Michele Pinchera}
\author[c]{Carmelo Sgr\`{o}}

\affil[a]{Istituto di Astrofisica e Planetologia Spaziali, Via Fosso del Cavaliere 100, I-00133 Rome, Italy}
\affil[b]{Universit\`{a} di Pisa, Dipartimento di Fisica Enrico Fermi, Largo B. Pontecorvo 3, I-56127 Pisa, Italy}
\affil[c]{Istituto Nazionale di Fisica Nucleare, Sezione di Pisa, Largo B. Pontecorvo 3, I-56127 Pisa, Italy}

\cftpagenumbersoff{figure}
\cftpagenumbersoff{table} 
\begin{document} 
\maketitle

\begin{abstract}
In this work, we measured the polarization properties of the X-rays emitted from the X-ray tubes, which were used during the calibration of the instrument onboard Imaging X-ray Polarimetry Explorer (IXPE). X-ray tubes are used as a source of unpolarized X-rays to calibrate the response of the gas pixel detectors to unpolarized radiation. However, even though the characteristic fluorescent emission lines are unpolarized, continuum bremsstrahlung emission can be polarized based on the geometry of the accelerated electrons and emitted photons. Hence, characterizing the contribution of polarized X-rays from bremsstrahlung emission is of interest, also for future measurements. We find that when accelerated electrons are parallel to the emitted photons, the bremsstrahlung emission is unpolarized, and when they are perpendicular, the polarization increases with energy, as expected from the theoretical predictions. A comparison with the theoretical predictions is also shown.
\end{abstract}


{\noindent\footnotesize\textbf{*}\linkable{ajay.ratheesh@inaf.it} }

\begin{spacing}{2}   

\section{Introduction}
\label{sect:intro}  
X-ray tubes are widely used in various scientific fields as a source of X-rays of known energy, depending on the material. When an electron beam strikes the anode, the characteristic X-ray lines generated in the X-ray tube can be used to calibrate X-ray detectors.
Such X-ray tubes were used to calibrate the response of the Gas Pixel Detector (GPD) onboard the Imaging X-ray Polarimetry Explorer (IXPE) to unpolarized photons \cite{soffitta_2021}  \cite{ixperef_2021,Muleri_2022}. Even though IXPE is equipped with radioactive sources for onboard in-orbit calibration purposes \cite{Ferrazzoli_2020}, a comprehensive and extensive calibration had to be conducted on the ground at different energies using X-ray tubes \cite{Muleri_2022,dimarco_2022}.
For this reason, even though the spectral properties of the X-rays arising from the X-ray tube are well known, their polarization properties are poorly studied because of a lack of suitable polarimeters, which is, in turn, necessary for the calibration of X-ray polarimeters like the GPD onboard IXPE. Hence it is crucial to constrain the polarization from these X-ray tubes while using them to calibrate polarimeters like the GPD onboard IXPE.\\

Measuring the response of the detector to unpolarized radiation is an essential procedure in the calibration of an X-ray polarimeter, because they are not exempt from spurious modulation that has to be carefully studied, calibrated, and fitered-out \cite{Rankin_2022}. The spurious modulation observed in a detector can be attributed to systematic effects inherent to the detector. Hence, the source of X-rays needs to be unpolarized to measure the spurious modulation independent of the source modulation. Even though fluorescent k-shell emission from X-ray tubes is unpolarized, the continuum Bremsstrahlung emission can be polarized based on the geometry of the X-ray tube \cite{Gluckstern_1953b}. The contribution of the partially polarized continuum to the unpolarized characteristic line must be addressed due to the finite energy resolution of the detector. Therefore, it is crucial to understand the contribution of polarization from the continuum emission, while using these X-ray tubes for measuring the response to unpolarized radiation of any X-ray polarimeter. The manufacturer does not provide the polarization properties of the X-ray tubes used for IXPE calibration since calculating the polarization of X-ray tubes theoretically from the first principle is difficult as it depends on the details of the geometry of the emission. Even if a procedure to decouple the intrinsic response of the instrument and the signal generated by the genuine partial polarization of the X-ray tube has been used for the calibration of detectors onboard IXPE \cite{Rankin_2022}, it is important to measure its intrinsic polarization as a cross-check and future reference. Moreover, in general, we notice that in X-ray Astronomy, after 60 years and around 50 space missions, a good reference of cross-calibrations has been achieved \cite{Sembay_2010, Madsen_2017} so that any new mission can benefit from ground facilities and a large sample of celestial sources sometimes observed simultaneously with more than one satellite. In the domain of polarimetry, IXPE is measuring the polarization of tens (potentially hundreds) of X-ray sources and is building the first catalog that will be the reference for any future experiment. It is, therefore, imperative to make a substantial effort to the best knowledge of the absolute values of the published results, including the level of systematics.\\

The experimental verification of X-ray tube polarization properties was only done now, as measuring the polarization in wide X-ray bands was difficult until the recent development of highly sensitive and wide-band photoelectric X-ray polarimeters. Moreover, measurements involving polarization for material science are currently performed in Synchrotron facilities. However, acquiring sufficient allocation of time from Synchrotron facilities for polarimetric calibrations of GPD like detectors, which demand prolonged exposure measurements, is impractical. For example, calibration of a detector unit of IXPE took 40 days of measurements \cite{Muleri_2022}. On the other hand, the constant availability of X-ray tubes renders them a convenient option for prolonged usage and enables adherence to the IXPE mission launch timeline.
\\

In this work, we outline the analysis of the measurements performed to understand the polarization properties of X-ray tubes. Section 2 describes the method of measuring X-ray polarization with photoelectric polarimeters, and section 3 gives a theoretical background of X-ray tubes and expected polarization from Bremsstrahlung. Section 4 shows the measurements and results, and in section 5, we conclude by discussing the observations.\\

\section{Measuring polarization with IXPE}
IXPE is the first dedicated mission with focusing optics in space to measure the polarization of X-rays \cite{ixperef_2021, soffitta_2021}. IXPE is a NASA Astrophysics Small Explorer (SMEX) mission developed in collaboration with the Italian Space Agency (ASI) and was launched on December 9, 2021.
The IXPE focal plane instrument, comprising three flight Detector Units (DU), each hosting a GPD and a spare unit, were developed by Istituto Nazionale di Astrofisica/Istituto di Astrofisica e Planetologia Spaziali (INAF-IAPS) in Rome and Istituto Nazionale di Fisica Nucleare (INFN) in Pisa. The GPD is a photoelectric polarimeter, which can image the photoelectron track and reconstruct the photoelectron emission direction and the absorption point.\\

IXPE opens a new window in X-ray astronomy by providing additional information on the polarization degree (PD) and polarization angle (PA) of X-rays along with the energy, time of arrival, and interaction point in the detector. With the addition of polarimetry, our understanding of the geometry and emission mechanism of various X-ray sources can be substantially improved. Specifically, polarization can help break the degeneracy of some geometrical and physical models developed based on spectroscopy alone.\\

The sensitivity of any X-ray polarimeter can be estimated by a quantity known as the minimum detectable polarization at 99$\%$ confidence ($MDP_{99}$). $MDP_{99}$ is defined as \cite{Weisskopf_2010,Strohmayer_2013} : 
\begin{equation}
MDP_{99} = \frac{4.29}{\mu} \times \frac{1}{R_S} \times \sqrt{\frac{R_S+R_B}{T}},
\end{equation}

Where $\mu$ is the modulation factor (measured without background), which is the response of a polarimeter to fully polarized X-rays, $T$ is the exposure time, $R_S$, and $R_B$ is the source and background count rate. When the $MDP_{99}$ is smaller, the sensitivity of a polarimeter is higher. Hence when $\mu$ is larger the sensitivity of the polarimeter is larger. \\

\begin{figure}[htb!]
\centering
\includegraphics[width=0.8\linewidth]{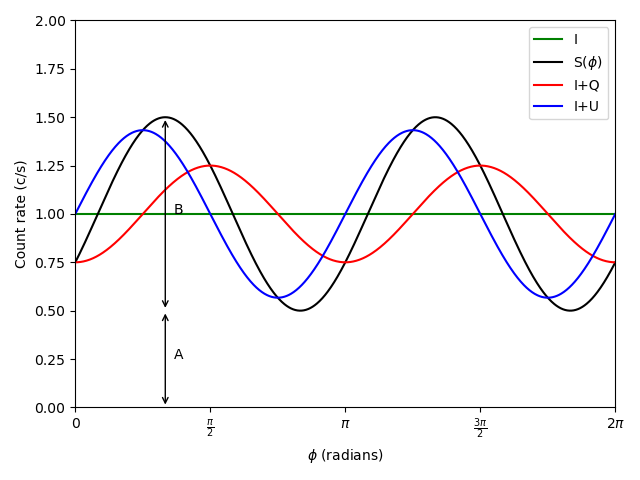}
\caption{An example modulation curve for linearly polarized light, decomposed into stokes parameters as labeled (for modulation amplitude, m = 0.5, and polarization angle ,$\phi_0$ = 60$^\circ$).}
\label{stokes_theo}
\end{figure}

In photoelectric absorption, the photoelectron's emission direction is parallel to the electric field of the incoming photon. Hence, the azimuthal distribution would be sinusoidally modulated depending on the degree and angle of polarization. Because the GPD is a photoelectric polarimeter, given the cross-section for K-shell electrons, the azimuthal angle distribution for photoelectrons (modulation curve) has a cosine dependence, well described by the following function  \cite{Strohmayer_2013}:
\begin{equation}
S(\phi) = A + B \cos 2 (\phi - \phi_0)
\end{equation} 
where $\phi_0$ is the polarization angle and the amplitude of modulation can then be described by \cite{Strohmayer_2013}: 
\begin{equation}
m = \frac{S_{max} - S_{min}}{S_{max} + S_{min}} = \frac{B}{2A+B} 
\end{equation} 
However, the amplitude of modulation is not the polarization degree of the observation and has to be normalized by the modulation factor of the detector to obtain the source polarization degree \cite{Strohmayer_2013}:
\begin{equation}
p = \frac{m}{\mu}. 
\end{equation} 
The polarization degree and angle can also be expressed in terms of the Stokes parameters. Stokes parameters are generally used to express the polarization of electromagnetic radiation (I, Q, U, V). The parameter I represents the intensity of the beam, Q represents the intensity in the $+90$ and $0$ directions. U represents the intensity in the $+45$ and $-45$ directions and V represents clockwise and anticlockwise circular polarization. The circular polarization arises when the electric field vector of X-rays oscillates in a circular pattern as it travels through space. However, in astronomical X-ray polarimetry, we do not consider circular polarization due to difficulty in measurement. If an X-ray polarimeter is designed to be sensitive only to linear polarization, the presence of circular polarization can potentially reduce the measured degree of linear polarization. An example of a modulation curve of polarized flux is shown in Figure. \ref{stokes_theo}. In terms of the modulation curve parameters, Stokes parameters can be expressed as\cite{Strohmayer_2013}:
\begin{equation}
I = A + B/2
\end{equation} 
\begin{equation}
Q = (B/2) \cos (2 \phi_0)
\end{equation} 
\begin{equation}
U = (B/2) \sin (2 \phi_0)
\end{equation} 
The modulation curve can be expressed in terms of Stokes parameters\cite{Strohmayer_2013} as:\\ \\
\begin{equation}
S(\phi) = I + Q \cos (2\phi) + U \sin (2\phi) 
\end{equation} 
Now the modulation amplitude ($m$) and angle ($\phi_0$) of modulation can be expressed in terms of the Stokes parameters as:
\begin{equation}
m = \frac{(Q^2 + U^2)^{0.5}}{I}
\end{equation} 
\begin{equation}
\phi_0 = 1/2  \tan^{-1} (U/Q)
\end{equation} 
However, the approach followed in this work and, in general, the case of IXPE is based on measuring the stokes parameters in an event-by-event approach to account for the subtraction of systematic effects \cite{Kislat_2015, Rankin_2022}. For each photon, the Stokes parameters are calculated as:

\begin{equation}
    q_i = 2 \; cos \; (2\phi_i)
\end{equation}
\begin{equation}
    u_i = 2 \; sin \; (2\phi_i)
\end{equation}
 where $\phi_i$ is the initial photo-electron direction, estimated by the photo-electron track reconstruction algorithm \cite{Bellazzini_2003a,Bellazzini_2003b}. 
And for a total number of N events
\begin{equation}
    q = \frac{\Sigma_i q_i}{N}
\end{equation}
\begin{equation}
    u = \frac{\Sigma_i u_i}{N}
\end{equation}
and the modulation and angle can be computed by 
\begin{equation}
    m = (q^2 + u^2)^{0.5}
\end{equation}
\begin{equation}
\phi_0 = 1/2  \tan^{-1} (u/q)
\end{equation}

\section{X-ray tubes and expected polarization}
X-ray tubes emit radiation with the following principle. Electrons are generated from a heated cathode and are accelerated towards an anode by a high voltage. The electrons hit the target anode, get decelerated, and produce X-rays due to bremsstrahlung emission. Bremsstrahlung emission is due to the deflection of the electrons by an atomic nucleus, and the energy lost by the electron is emitted as a photon. The maximum energy of the bremsstrahlung emission depends on the peak accelerating potential of the X-ray tube, and the bremsstrahlung peaks at approximately 1/3$^{rd}$ of the maximum energy \cite{Agarwal_2013}. The intensity of the bremsstrahlung emission is directly proportional to the atomic number of the target material and the charge of the particle and inversely proportional to the mass of the particle, in our case, the electron. The spectrum of bremsstrahlung emission in an X-ray tube can be described by Kramers law \cite{Kramers_1923}:
\begin{equation}
    d \; I(\lambda) \; = \; K \; \biggl(\frac{\lambda}{\lambda_{min}} - 1\biggr) \; \frac{1}{\lambda^2} \; d\lambda
\end{equation}
where K is a constant depending on the electron beam current, $\lambda$ is the wavelength of the emitted photons, and, $\lambda_{min}$ is the minimum wavelength of the emitted photons from an X-ray tube for a given applied voltage ($V$), given by the Duane–Hunt law \cite{Duane_hunt_1915}:
\begin{equation}
    \lambda_{min} \; = \; \frac{hc}{eV}
\end{equation}
Other than the bremsstrahlung emission, the X-ray tubes also emit characteristic fluorescent X-rays when the accelerated electrons hit the target material and kick an electron from the K-shell, and get refilled by an outer shell electron. The energy of the discrete fluorescent lines depends on the binding energy of the electron in the target material. About one percent of the energy of the accelerated electrons is radiated through X-rays. The high voltage applied to the X-ray tube will change the spectral shape, while the applied current will change the intensity or normalization. \\

\subsection{Theoretical prediction of polarization from bremsstrahlung emission}
Theoretical calculations of the cross-section and polarization from the bremsstrahlung emission were already published in the middle of 20$^{th}$ century \cite{Sommerfeld_1931,Heitler_1933,Kirkpatrick_1945,Gluckstern_1953a,Gluckstern_1953b}. It was found that the bremsstrahlung emission is supposed to be partially polarized. The primary step in determining the polarization is to determine the cross-section. The bremsstrahlung cross-section gives the probability of an electron transiting from one to another state with the emission of a photon. The electron-nucleus collisions can also be elastic, and hence the probability that a photon will be emitted depends on the cross-section of the bremsstrahlung, which is 137 times smaller than the cross-section of electron-nucleus elastic scattering. The differential cross-section ($d\sigma$) as a function of the photon energy for an electron emerging through the solid angle $d\Omega$ in the non-relativistic limit, which is the case for electron kinetic energies smaller than 10 keV, can be expressed as:
\begin{equation}
    d\sigma \; = \; \biggl(\frac{Z^2 e^6}{\pi^2}\biggr) \; \biggl(\frac{p}{p_0}\biggr) \; \biggl( \frac{dk}{kq^4} \biggr) \; d\Omega \; d\Omega_0 \; (p_l - p_{0l})^2
\end{equation}
where Z is the atomic number or nuclear charge, $e$ is the elementary charge, $k$ is the energy of the emitted photon, $q$ is the momentum transferred to the nucleus, $p$ and $p_0$ are the final and initial momentum of the electron, and $p_l$ and $p_{0l}$ are the components of momentum of the electron in the direction of the polarization. Hence the differential cross section for a specified photon energy further depends on the atomic number of the material, the direction of the initial and emerging electron, and the momentum of the initial and emerging electron.\\

Given the energy and momentum of the initial and final electron, along with the photon energy and momentum, linear polarization can be calculated by 
\begin{equation}
\label{brehm_theore_eqn3}
    P \; = \; \frac{d\sigma_{\rom{3}} - d\sigma_{\rom{2}} }{d\sigma_{\rom{3}} + d\sigma_{\rom{2}} } 
\end{equation}
Here, $d\sigma_{\rom{3}}$ is the differential cross-section of the polarization vector perpendicular to the plane of scattering, containing the initial electron momentum and photon momentum, and $d\sigma_{\rom{2}}$ is the differential cross-section of the polarization vector in the plane of scattering and parallel to $p$ \cite{Gluckstern_1953b}. It is to be noted that the notations used here are the same as that of Gluckstern et al. 1953\cite{Gluckstern_1953b} . The expressions for the $d\sigma_{\rom{2}}$ and $d\sigma_{\rom{3}}$:

\begin{equation}
\label{brehm_theore_eqn1}
\begin{aligned}
    d\sigma_{\rom{2}} \; = \: \frac{Z^2 e^6}{8\pi} \frac{dk}{k} \; \frac{p}{p_0} \; d\Omega_0 \; 
     \bigg\{
    \frac{8m^2 sin^2\theta_0 (2E^2_0+m^2)}{p_0^2 \Delta_0^2}
    -\frac{5E_0^2+2EE_0+5m^2}{p_0^2\Delta_0^2} 
    -\frac{p_0^2-k^2}{T^2\Delta_0^2}
    +\frac{2(E+E_0)}{p_0^2\Delta_0}\\
  + (\frac{L}{pp_0})\big[ \frac{4E_0m^2sin^2\theta_0(3km^2-p_0^2E)}{p_0^2\Delta_0^4}
  +\frac{2E_0^2(E_0^2+E^2)-m^2(9E_0^2-4EE_0+E^2)+2m^4}{p_0^2\Delta_0^2} + k\frac{E_0^2+EE_0}{p_0^2\Delta_0} \big] \\
 (\frac{\epsilon^T}{pT}) \big[ \frac{4m^2}{\Delta_0^2}-\frac{7k}{\Delta_0}-k\frac{p_0^2-k^2}{T^2\Delta_0}-4 \big] -\frac{4\epsilon}{p\Delta_0} 
 +(\frac{1}{p_0^2sin^2\theta_0^2}) \big[ (\frac{2L}{pp_0})(2E_0^2-EE_0-m^2-\frac{m^2k}{\Delta_0}) \\
 - \frac{4\epsilon^T(\Delta_0-E)^2}{pT} -\frac{2\epsilon(\Delta_0-E)}{p} \big] \bigg\}
 \end{aligned}
 \end{equation}

\begin{equation}
\label{brehm_theore_eqn2}
\begin{aligned}
    d\sigma_{\rom{3}} \; = \: \frac{Z^2 e^6}{8\pi} \frac{dk}{k} \; \frac{p}{p_0} \; d\Omega_0 \; 
     \bigg\{
-\frac{5E_0^2+2EE_0+m^2}{p_0^2\Delta_0^2} -\frac{p_0^2-k^2}{T^2\Delta_0^2}-\frac{2k}{p_0^2\Delta_0} \\
+ (\frac{L}{pp_0})\big[ \frac{2E_0^2(E_0^2+E^2)-m^2(5E_0^2-2EE_0+E^2)}{p_0^2\Delta_0^2} + \frac{k(E_0^2+EE_0-2m^2)}{p_0^2\Delta_0} \big]\\
+(\frac{\epsilon^T}{pT}) \big[ \frac{k}{\Delta_0} - \frac{k(p_0^2-k^2)}{T^2\Delta_0} + 4 \big]
-(\frac{1}{p_0^2sin^2\theta_0^2}) \big[ (\frac{2L}{pp_0})(2E_0^2-EE_0-m^2-\frac{m^2k}{\Delta_0})\\
 - \frac{4\epsilon^T(\Delta_0-E)^2}{pT} -\frac{2\epsilon(\Delta_0-E)}{p} \big] \bigg\}
\end{aligned}
\end{equation}

Where,
\begin{equation}
\Delta_0 = E_0 - p_0 cos\theta_0,   
\end{equation} 
\begin{equation}
T = p_0^2+k^2-2p_0kcos\theta_0,   
\end{equation}
\begin{equation}
L = ln \bigg( \frac{EE_0-m^2+pp_0}{EE_0-m^2-pp_0} \bigg),
\end{equation}
\begin{equation}
\epsilon = ln \bigg( \frac{E+p}{E-p} \bigg),
\end{equation}
\begin{equation}
\epsilon^T = ln \bigg( \frac{T+p}{T-p} \bigg),
\end{equation}


and $\theta_0$ is the angle between $p_0$ and $k$. $E$ and $E_0$ represents the initial and final energy of the incoming electron. The constants $\hbar$ and c are taken to be 1. The effects of shielding are complicated. However, they can be approximated by replacing $k^2$ by $k^2+\alpha^2p_0^2\Delta_0^{-2}$ in the logarithmically divergent terms $L$ and $\epsilon^T$. Hence, 
\begin{equation}
    L = ln \; \bigg( \; \frac{(EE_0-m^2+pp_0)^2}{m^2k^2+m^2\alpha^2p_0^2\Delta_0^{-2}} \; \bigg),
\end{equation}

\begin{equation}
    \epsilon^T \; = \; \frac{1}{2} ln \bigg( \frac{(T+p)^4}{4k^2\Delta_0^2+4\alpha^2p_0^2} \bigg),
\end{equation}
where,
\begin{equation}
\alpha = Z^\frac{1}{3}\frac{m}{108}
\end{equation}

\begin{figure}[htb!]
\centering
\includegraphics[width=0.8\linewidth]{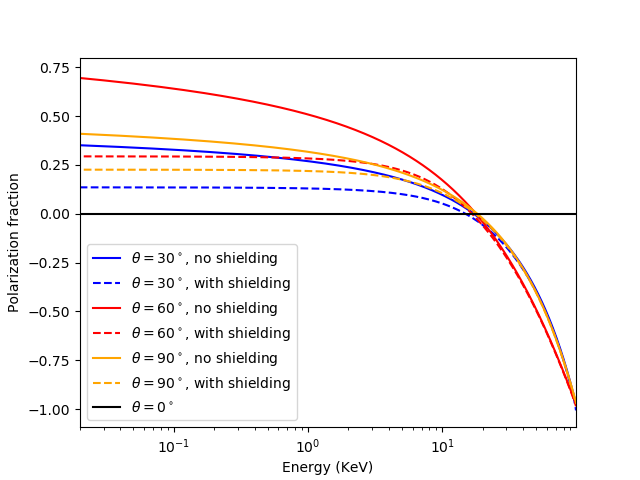}
\caption{Polarization degree of the outgoing photons as a function of photon energy for incident electron energies of 0.1 MeV for Z=13 (Aluminium) for different values of the angle between incoming electrons and outgoing photons ($\theta_0$) as well as with and without shielding effects.} 
\label{xraytube_theopred_pol}
\end{figure}

\begin{figure}[htb!]
\centering
\includegraphics[width=0.8\linewidth]{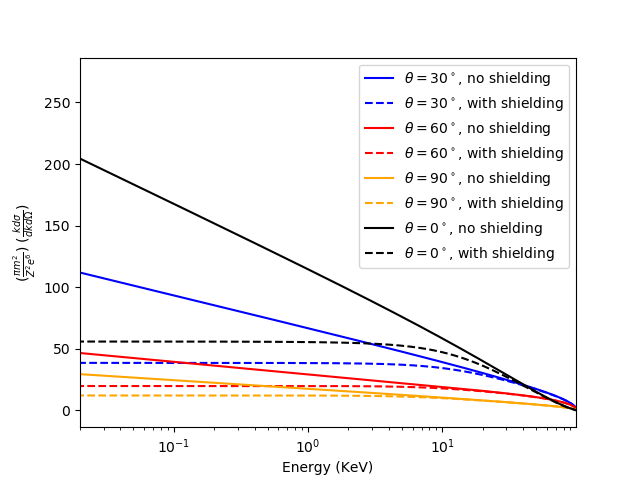}
\caption{The cross-section of the outgoing photons as a function of photon energy for incident electron energies of 0.1 MeV for Z=13 (Aluminium) for different values of the angle between incoming electrons and outgoing photons ($\theta_0$) as well as with and without shielding effects.}.
\label{xraytube_theopred_spec}
\end{figure}

Figures \ref{xraytube_theopred_pol} and \ref{xraytube_theopred_spec} shows the degree of polarization and cross-section with respect to photon energy for a beam of electron of energy 0.1 MeV in aluminum \cite{Gluckstern_1953b}. It is seen that the polarization is high around the lower and upper ends of the photon spectrum. A jump in the polarization angle by 90$\degree$ is seen as a change in the sign of the polarization degree at medium energies of approximately 20 keV. The shielding effect decreases the polarization degree, especially at lower energies. 



\section{Measurements and Results}
\begin{figure}[htb!]
\centering
\includegraphics[width=0.8\linewidth]{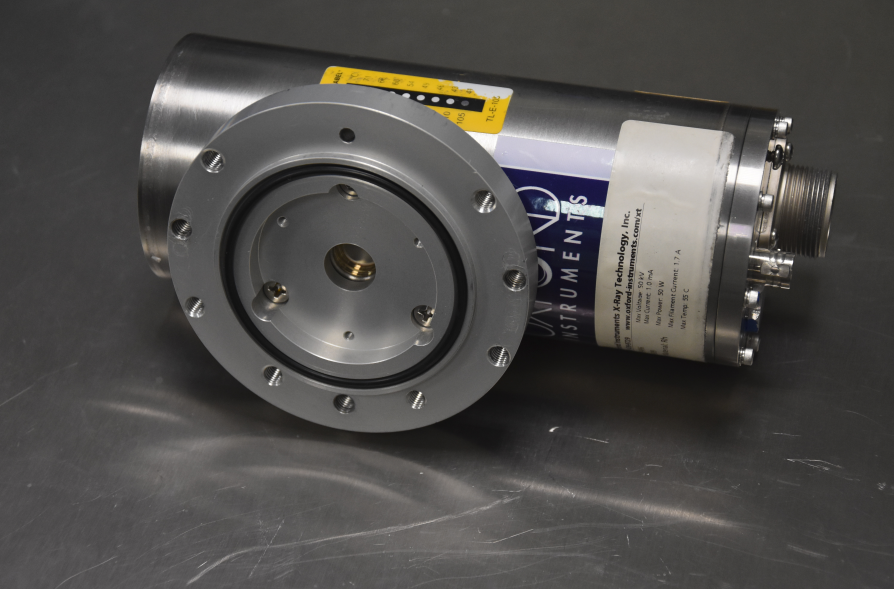}
\caption{Photograph of the Iron X-ray tube used as a perpendicular X-ray tube in this work.}
\label{Fe_tube_pc}
\end{figure}

\begin{figure}[htb!]
\centering
\includegraphics[width=0.8\linewidth]{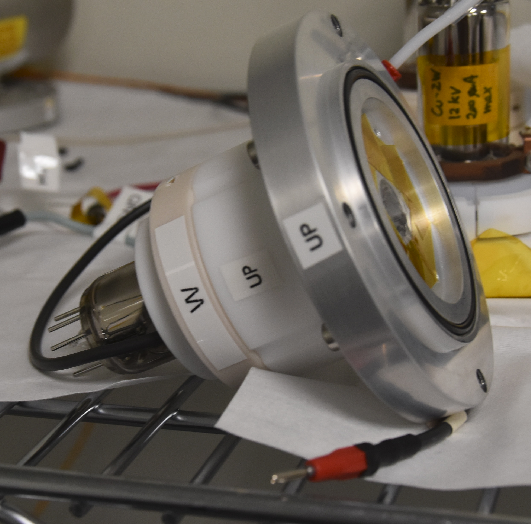}
\caption{Photograph of the tungsten X-ray tube used as a parallel X-ray tube in this work. }
\label{W_tube_pc}
\end{figure}

\begin{figure}[htb!]
\centering
\includegraphics[width=0.8\linewidth]{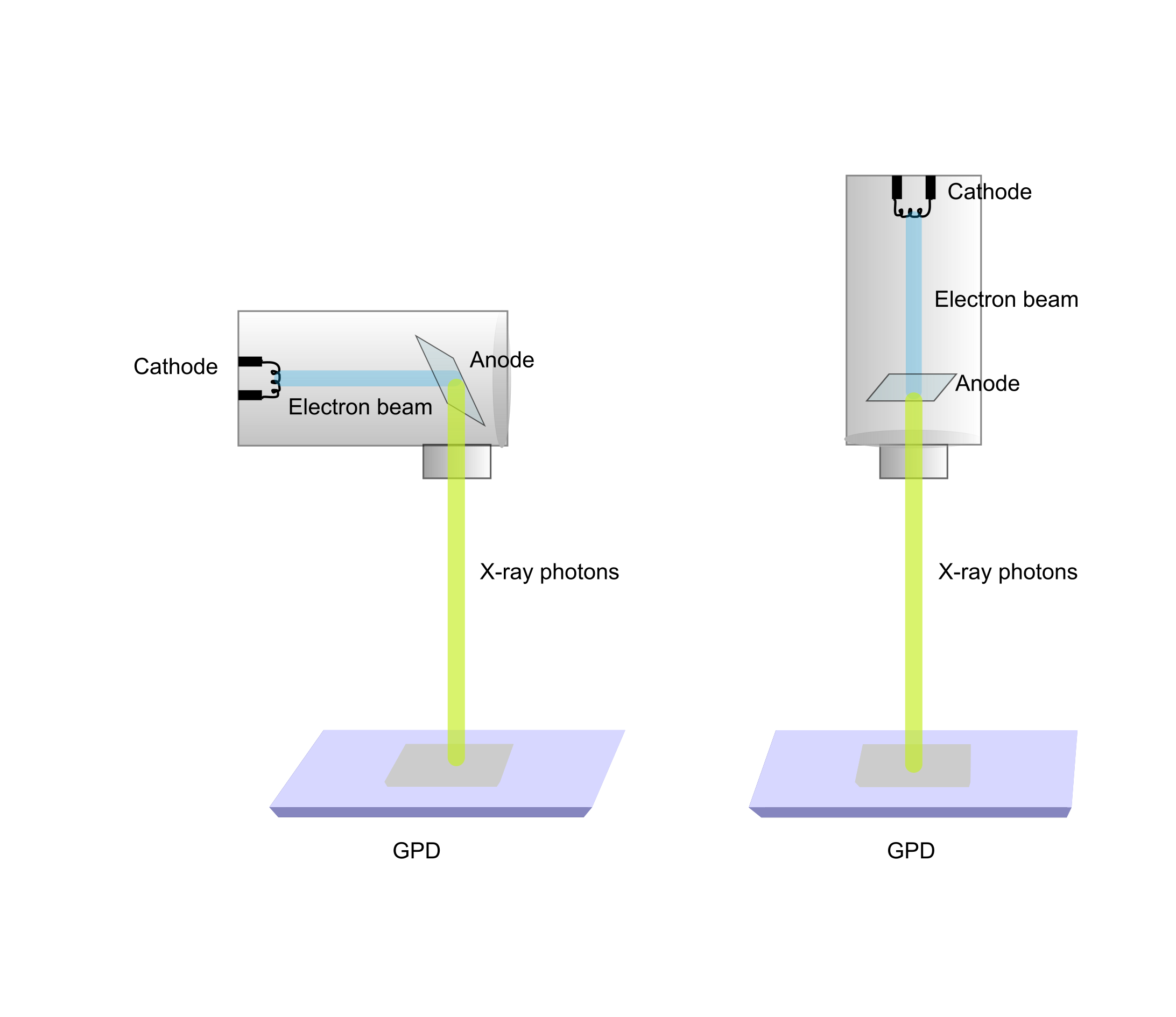}
\caption{A rough sketch of the experimental geometrical configuration of perpendicular (left) and parallel (right) X-ray tubes with respect to the gas pixel detector (GPD). In the former, the photon beam is perpendicular to the electron beam while in the later both are parallel.}
\label{xraytube_geometry}
\end{figure}

We now present the polarization measured for the X-ray tubes used to calibrate the GPDs onboard IXPE. There are different X-ray tubes used for the calibration of IXPE \cite{Muleri_2022}, and are primarily two kinds, right angle X-ray tubes from Oxford series 5000 (Fig. \ref{Fe_tube_pc}) and head-on X-ray tubes like the Calcium and Tungsten anode ones from Hamamatsu, model N1335 (Fig. \ref{W_tube_pc}). The experimental setup of the measurements are the same as what is mentioned in \cite{Muleri_2022, dimarco_2022}. The source and detector were mounted to the instrument calibration equipment (ICE), an equipment specifically designed for calibrating IXPE DUs \cite{Muleri_2022}. This equipment encompasses the mechanical frameworks employed to secure the detector and sources, calibration sources, stages for aligning the detector and source beam, test detectors and their corresponding mechanical assembly (Figure 3 in \cite{Muleri_2022}). The alignment process involves two stages. The first stage, named ALIGN, aims to synchronize the source to the detector. The second stage, referred to as MEAS, facilitates movement of the detector along the detector plane, orthogonal shifts, azimuthal rotations, and tilting of the detector to align the beam with the detector plane. In the case of right-angle X-ray tubes, the generated X-ray photons are perpendicular to the accelerated beam of electrons. In contrast, in the case of head-on X-ray tubes, the X-ray photons are parallel to the accelerated beam of electrons. During the calibration of the IXPE Instrument, X-ray filters were used to increase the ratio between the fluorescence and the continuum emission, as only the former component is unpolarized and of interest for calibration. For our study, such filters were not used, as our primary aim is to measure the polarization of the bremsstrahlung continuum. For measuring the polarization we used an IXPE flight detector unit called as DU FM1 (Detector Unit Flight Model 1), which is the spare unit of the IXPE Instrument and was fully calibrated as the other flight units. The measurements were undertaken in the ICE after mounting DU FM1 and the different X-ray tubes on the ICE. Figure \ref{xraytube_geometry} shows the sketch of the geometrical configuration of the X-ray tubes and the detector.



Measurements were conducted at two arbitrary angles ($\epsilon1$ and $\epsilon2$) that are mutually perpendicular to each other in the detector plane ((ASIC X-Y Axis)) and with respect to the laboratory frame of reference. The measurements at two angles are utilized to ensure that the polarization that is measured originates from the source itself, as the polarization vector of the source will be rotated with respect to the reference frame. If the polarization indeed arises from the source and not from any detector systematics, then the measured polarization angle must undergo a 90$^\circ$ rotation upon rotating the source.
Table \ref{table_xraytube_measurements} outlines the details of the X-ray tubes and measurements used for this work. In the case of the Fe and W X-ray tubes, the high voltage was chosen deliberately to be below the K-shell electron binding energy to avoid the bright K fluorescence line from the material of the tube target.\\

For the analysis of the measurements, the spurious modulation due to the detector systematics has to be taken care of and subtracted carefully as in the case of celestial sources. A method to subtract the contribution from the detector on an event-by-event basis was developed for the IXPE flight detectors \cite{Rankin_2022}. We followed this method for subtracting the spurious modulation. A spurious modulation database developed for calibrating IXPE\cite{Rankin_2022}, which contains the spurious modulation as a function of spatial and energy bins, was used to subtract the detector contribution while extracting the polarization from the source.\\

The data calibrated for spurious modulation were then used for further analysis. We used all events within the circular spatial region of radius 2 mm from the center of the detector, as this is within the regions of the detector that is calibrated with the highest sensitivity. The data from the selected spatial region were then grouped into four different energy bins, with larger bin sizes at energy ranges with lower count-rate, to allow sufficient statistics for measuring the polarization. The normalized and spurious modulation corrected Stokes parameters from all events in each bin are then added to get the Stokes parameters and modulation amplitude of the total measurement. The total Stokes parameters and the modulation degree are then divided by the modulation factor of DU FM1 in that particular energy bin to get the final Stokes parameters and the polarization from the X-ray tubes \cite{dimarco_2022}. The modulation factor in each energy bin is weighted by the spectral counts in that bin.\\

In the case of the measurements with Rh and Ca X-ray tube, we could correct the gain variation during the measurement caused by charging of the gas electron multiplier (GEM), which multiplies the primary charges before collection \cite{Baldini_2021}, using the fluorescent line at the 2.7 and 3.69 keV. All the events were re-scaled by a correction factor: the ratio of the fluorescent line energy and the observed Gaussian line peak energy at different times. For the case of Fe and W X-ray tubes, there were no fluorescent lines as the applied high voltage was lower than the line energy; hence, these measurements are slightly affected by charging. This means that the measured energy in those measurements can have uncertainties of upto 10$\%$. 

\begin{figure}[htbp]
\centering
\includegraphics[width=0.8\linewidth]{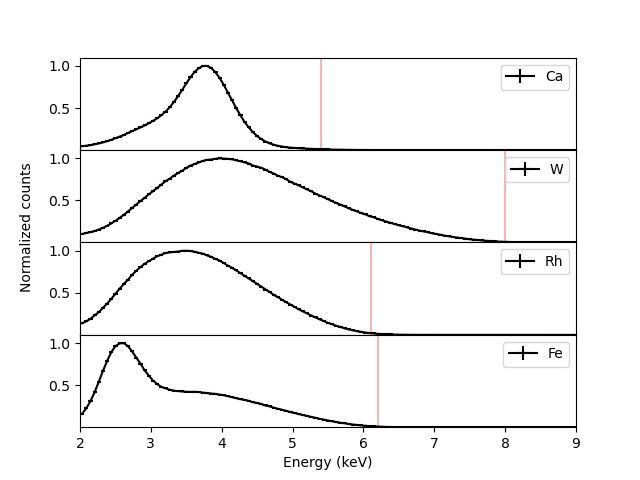}
\caption{The spectrum of all the X-ray tubes used in this work as seen in the GPD for $\epsilon1$. The spectrum of each X-ray tube has been normalized by the corresponding peak of the spectrum. The vertical lines indicate the maximum possible energy of the X-rays from the tube corresponding to the applied voltage. For W and Rh tubes the spectrum has only continuum Bremsstrahlung, as the applied voltage is below the characteristic line energy. The spectrum of $\epsilon2$ is identical to that of $\epsilon1$, and thus it has not been presented to avoid redundancy.}
\label{xraytube_spec}
\end{figure}

\begin{figure}[htbp]
\centering
\includegraphics[width=1.2\linewidth]{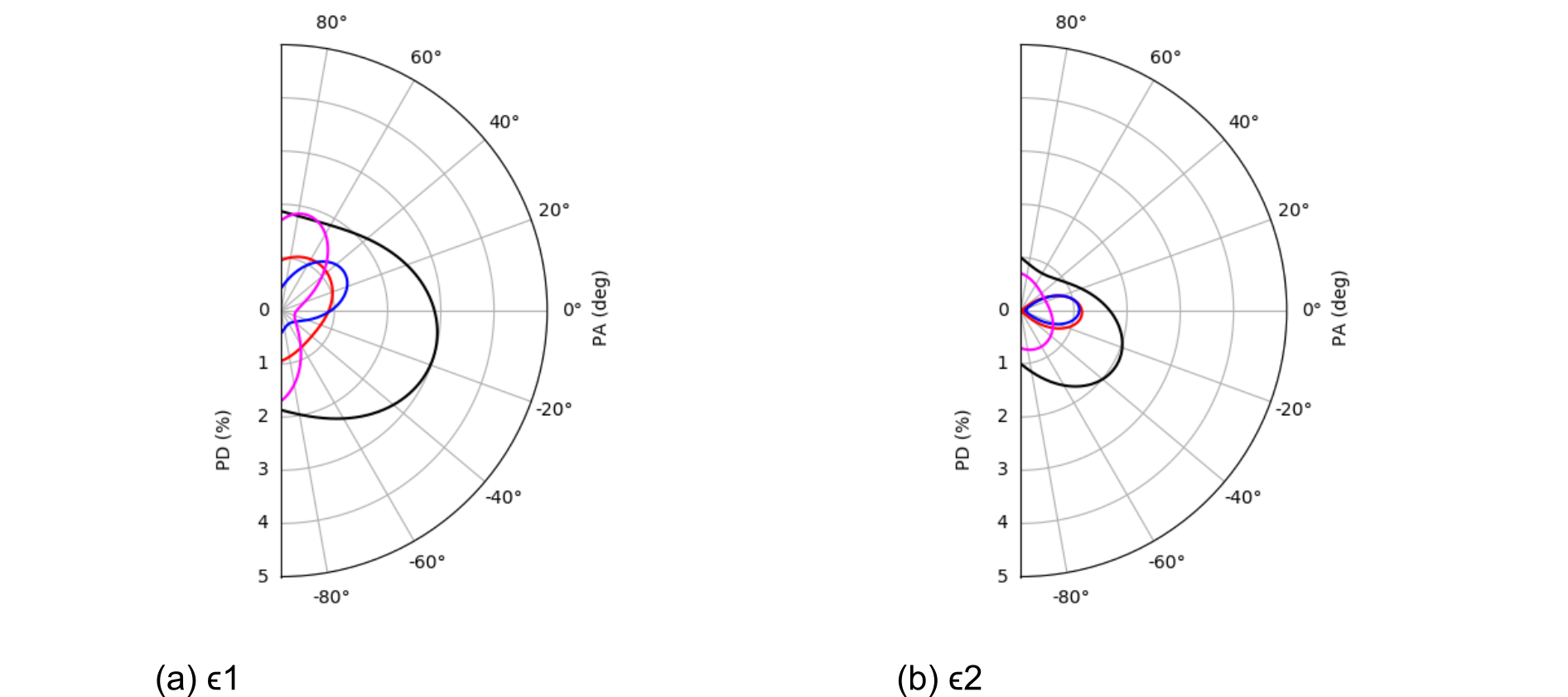}
\caption{Polar plots of the polarization degree (PD) and polarization angle (PA) for the measurements of Ca X-ray tube at angles of $\epsilon1$ and $\epsilon2$. Contours indicates 99.7\% confidence levels. Colors black, red, blue, and magenta indicate energy ranges 2.0-3.0 keV, 3.0-3.7 keV, 3.7-4.2 keV, and 4.2-8.0 keV. These contours indicate that there are no significant polarization measurements but only constraints of upper-limits.}
\label{Ca_xraytube_contour}
\end{figure}

\begin{figure}[htbp]
\centering
\includegraphics[width=0.8\linewidth]{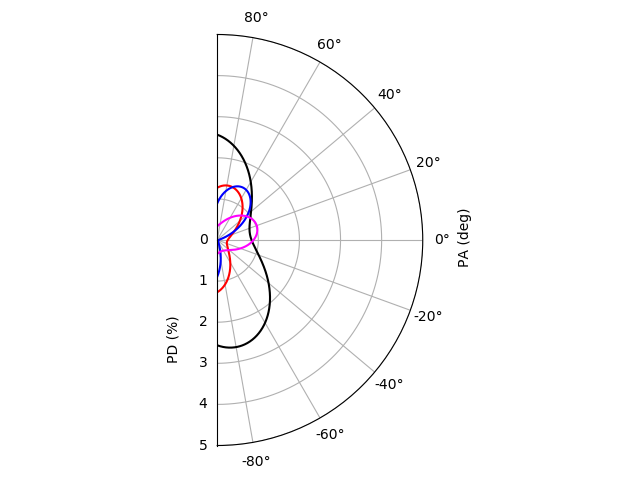}
\caption{Same as Figure \ref{Ca_xraytube_contour}, but for W X-ray tube and only for the angle of $\epsilon1$. Contours indicates 99.7\% confidence levels. Colors black, red, blue, and magenta indicate energy ranges 2.0-3.25 keV, 3.25-4.25 keV, 4.25-5.25 keV, and 5.25-8.0 keV. } 
\label{W_xraytube_contour}
\end{figure}

\begin{figure}[htbp]
\centering
\includegraphics[width=1.2\linewidth]{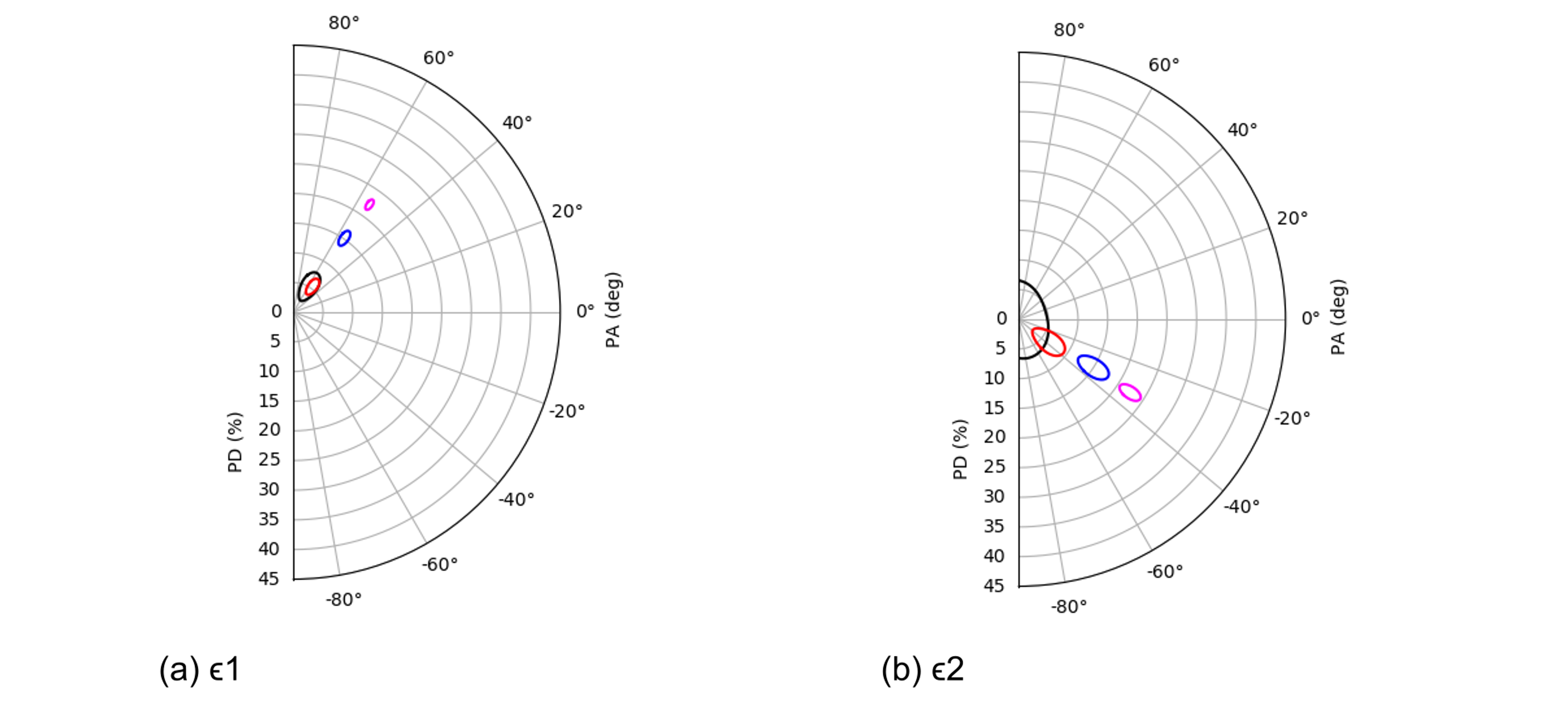}
\caption{ Polar plots of the polarization degree (PD) and polarization angle (PA) for the measurements of Rh X-ray tube at angles of $\epsilon1$ and $\epsilon2$.  Contours indicates 99.7\% confidence levels. Black, red, blue, and magenta indicate energy ranges of 2.0-2.6 keV, 2.6-3.2 keV, 3.2-4.0 keV, and 4.0-7.0 keV. The increase in polarization with respect to energy can be seen clearly. The 90$^\circ$ phase shift observed in the polarization angle (PA) between $\epsilon1$ and $\epsilon2$ confirms that the PA measurement undergoes a 90$^\circ$ rotation when the source is rotated by 90$^
\circ$, and hence confirming the source origin of polarization than detector systematics.}
\label{Rh_xraytube_contour}
\end{figure}

\begin{figure}[htbp]
\centering
\includegraphics[width=1.2\linewidth]{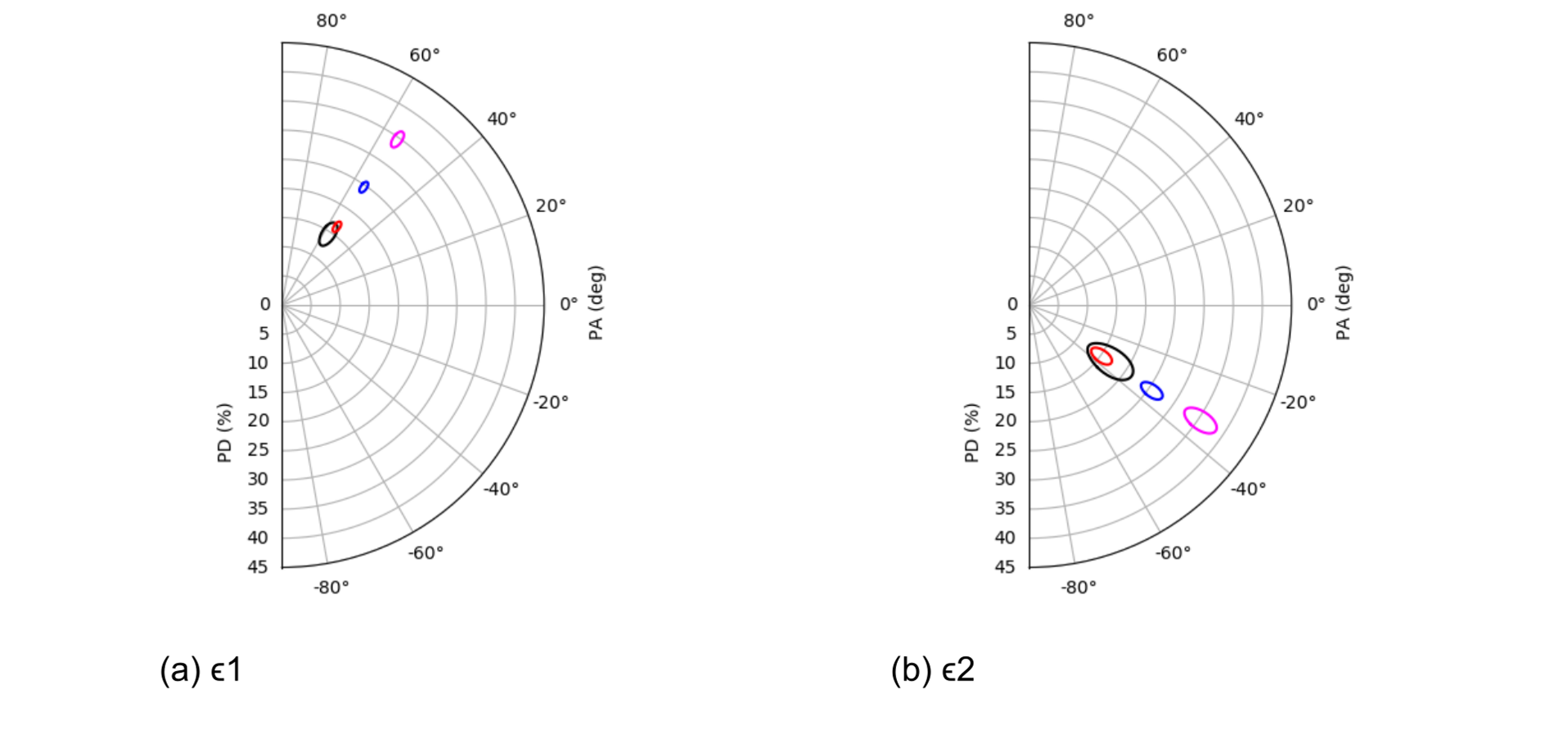}
\caption{ Same as Figure \ref{Rh_xraytube_contour}, but for Fe X-ray tube. Contours indicates 99.7\% confidence levels. Black, red, blue, and magenta indicate energy ranges of 2.0-3.0 keV, 3.0-4.0 keV, 4.0-5.0 keV, and 5.0-8.0 keV.}
\label{Fe_xraytube_contour}
\end{figure}

Figure \ref{xraytube_spec} shows the spectra, while Figures \ref{Ca_xraytube_contour}, \ref{W_xraytube_contour}, \ref{Rh_xraytube_contour}, and \ref{Fe_xraytube_contour} shows the contours of polarization degree and angle. The corresponding data for these measurements are noted down in Tables \ref{table_Ca}, \ref{table_W}, \ref{table_Rh},  and \ref{table_Fe}. For plotting the spectrum, the 2-8 keV is divided into 100 energy bins and normalized by the peak of the spectrum. \\

The polarization degree was seen to increase with energy for the right angle X-ray tubes (Fe and Rh), while for head-on X-ray tubes (Ca and W), the X-rays are unpolarized, while the measured polarization angle in both cases seems to be parallel to the direction of the incoming electrons, as what was expected in the energy range of the GPD. In the case of Rh and Fe X-ray tube, the polarization increases from around $5\%$ to $22\%$ and around $15\%$ to $35\%$ across the IXPE energy band of 2 to 8 keV (see Table \ref{table_Rh} and Table \ref{table_Fe}). Even though the increasing trend of polarization is as expected, the absolute values of the measured polarization are lower than what is expected for an ideal case of single scattering of the electron in a crystal. However, in our case, the mean free path of the electron with an energy of few keV in Rh and Fe (of the order of nm) would be much less than the thickness of the anode target (a few tens of microns), increasing the probability of multiple scattering of the single electron giving rise to multiple Bremsstrahlung photons. This effect would, in turn, dilute and decrease the polarization degree. It was previously seen that even for very thin targets and at larger energies the measured polarization was smaller than the theoretically expected value \cite{Kuckuck_1973}. In the case of Ca and W X-ray tubes, we obtained an upper limit closer to $1\%$ or lesser (see Table \ref{table_Ca} and Table \ref{table_W}). At some energies, we have a measurement slightly above the MDP, but smaller than $1\%$, and it could be due to some unknown systematic like the multiple scattering mentioned above. The shift in the polarization angle for $\epsilon2$ by 90$\degree$, with respect to $\epsilon1$ further indicates that the polarization is coming from the source rather than a detector systematic like the spurious modulation (which indeed we subtract). If the polarization was a detector effect, the polarization angle should not have rotated by 90$\degree$ while rotating the illuminating source. It can also be seen that the highest energy in the photon spectrum matches with the applied high voltage of the tube.\\

\begin{figure}[htb!]
\centering

\includegraphics[width=\linewidth, height=0.8\linewidth]{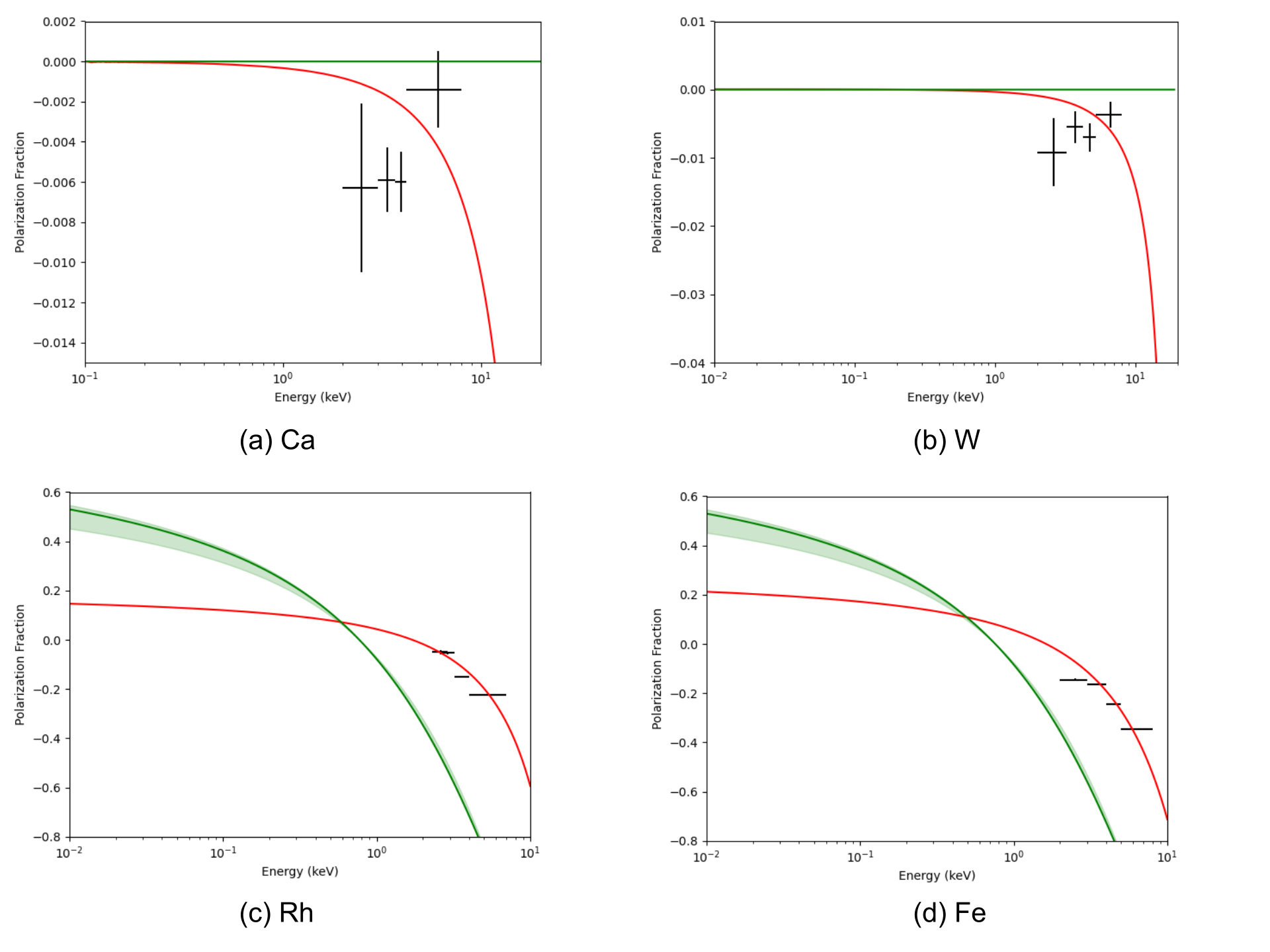}
\caption{The polarization fraction as a function of the photon energy for all the X-ray tubes (at $\epsilon1$). The black points show the measured polarization. The solid green line indicates the expected value from the theory, while the shaded green region in the case of Fe and Rh tubes shows the range due to the non-zero opening angle subtended by the crystal on the detector. The red line indicates the values of the best fit. The negative polarization fraction means that the polarization angle is parallel to the electron direction, while the positive indicates the polarization angle perpendicular to the incoming electron and outgoing photon. }
\label{fit_theoretical_func}
\end{figure}

The polarization measured at $\epsilon1$ is fitted with the theoretical function (Equations \ref{brehm_theore_eqn3} to \ref{brehm_theore_eqn2}), allowing the applied voltage (V) and angle between the incoming electron and the outgoing photon ($\theta$) left for the fit to determine. The results of the fit are shown in Fig. \ref{fit_theoretical_func}. The solid red line is the fit to the data, while the green lines indicate the expected theoretical value based on V and $\theta$. For the case of the perpendicular X-ray tubes, the opening angle of the anode crystal with respect to the aperture towards the detector is 11$^\circ$, and hence to account for it within the theoretical estimate, we shade the space around the line assuming an error of 11$^\circ$ from the tube. However, the deviation is well beyond this limit. Table. \ref{fit_theoretical_func_table} outlines the fit results for all the X-ray tubes. In the case of the parallel X-ray tubes, the fitted $\theta$ is closer to zero, while the fitted V is 3-4 times higher than the real applied voltage. This is mostly because of the non-zero values of polarization. In the case of the perpendicular X-ray tubes, both the $\theta$ and V are different from the real applied values. V is higher by a factor of 2, while $\theta$ is smaller by a factor of 3. This mismatch of values could be due to the multiple electron scattering effects. \\

\section{Conclusions}
In this work, we calculated the polarization of the continuum Bremsstrahlung emission from X-ray tubes used to calibrate the GPD onboard IXPE. The theoretical predictions of polarization from Bremsstrahlung depend on the angle between the incoming electron and the emitted photon ($\theta_{0}$). When $\theta_{0}$=0$\degree$, the polarization is supposed to be constant at $0\degree$, and when $\theta_{0}$=90$\degree$, the polarization is supposed to decrease until a certain energy and then increase. In the case of X-ray tubes, the $\theta_{0}$ depends on the geometrical configuration of the X-ray tube. We used 2 X-ray tubes where $\theta_{0}$=0$\degree$ and 2 X-ray tubes where $\theta_{0}$=90$\degree$. 
For perpendicular X-ray tubes, we find that the measured polarization degrees are significantly lower than the ones predicted by theory, despite a good match between the observed polarization angle and the overall trend of the polarization degree with energy in the 2-8 keV range. This mismatch of the absolute value could be due to the multiple electron scattering in the anode crystal, which is relevant in 2-8 keV. This could not be tested as the manufacturer does not provide the dimensions of the anode crystal. Nonetheless, this scenario is plausible, considering that the mean free path of electrons with energies in the range of a few keVs in Ca, W, Rh and Fe is of the order of nanometers, whereas the thickness of the anode could exceed micrometers. For parallel X-ray tubes, we obtained upper-limits consistent with the predictions. \\

Our work experimentally shows the differences in the continuum polarization of two different configurations of X-ray tubes. The fact that the general trend of polarization degree and polarization angle is coherent with what is expected is positive. However, our work aimed to quantitatively fix the values of practical implementation to be used to calibrate detectors of IXPE or any other detector. These measurements can now form a base while using these X-ray tubes in the future to understand the response of any X-ray polarimeter to unpolarized X-rays.\\
\subsection* {Acknowledgments}
The Imaging X-ray Polarimetry Explorer (IXPE) is a joint US and Italian mission. The US contribution is supported by the National Aeronautics and Space Administration (NASA) and led and managed by its Marshall Space Flight Center (MSFC), with industry partner Ball Aerospace (contract NNM15AA18C). The Italian contribution is supported by the Italian Space Agency (Agenzia Spaziale Italiana, ASI) through contract ASI-OHBI-2017-12-I.0, agreements ASI-INAF-2017-12-H0 and ASI-INFN-2017.13-H0, and its Space Science Data Center (SSDC) with agreements ASI-INAF-2022-14-HH.0 and ASI-INFN 2021-43-HH.0, and by the Istituto Nazionale di Astrofisica (INAF) and the Istituto Nazionale di Fisica Nucleare (INFN) in Italy. This research used data products provided by the IXPE Team (MSFC, SSDC, INAF, and INFN) and distributed with additional software tools by the High-Energy Astrophysics Science Archive Research Center (HEASARC), at NASA Goddard Space Flight Center (GSFC).



\bibliography{report}   
\bibliographystyle{spiejour}

{\bf Ajay Ratheesh} is a Post-Doctoral researcher affiliated with INAF-IAPS, situated in Rome. In 2021, he earned his PhD in Astronomy and Astrophysics through the joint PhD program offered by Sapienza University of Rome, University of Rome "Tor Vergata", and INAF in Italy. Currently, he is a member of the IXPE collaboration. \\
\noindent Biographies and photographs of the other authors are not available.
\clearpage

\section{Tables}

\begin{table}[h]
\caption{The table outlining the details of the measurement configurations used in this work.}
\label{table_xraytube_measurements}
\centering   
\renewcommand{\arraystretch}{2.0}
\begin{tabular}{|c|c|c|c|c|c|} 
\hline\hline  
X-ray tube & Anode & Fluorescent line energy & Source configuration & Angle\\
\hline\hline
Hamamatsu, model N1335 & Ca & 3.69 & 5.4 kV, 0.003 mA & $\epsilon1$, $\epsilon2$ \\
Hamamatsu & W & 8.34 & 8.0 kV, $<$ 0.001 mA  & $\epsilon1$ \\
Oxford series  & Fe & 6.40 & 6.1 kV, 0.003 mA & $\epsilon1$, $\epsilon2$ \\
Oxford series 5000 & Rh & 2.70 & 6.2 kV, 0.001 mA & $\epsilon1$, $\epsilon2$ \\
\hline
\hline
\end{tabular}
\end{table}

\begin{table}[h]
\caption{Polarization results from the Ca X-ray tube for $\epsilon1$ and $\epsilon2$. }
\label{table_Ca}
\centering   
\renewcommand{\arraystretch}{1.0}
\begin{tabular}{|c|c|c|c|c|c| } 
\hline\hline
Energy (keV) & Q/I ($\%$) & U/I ($\%$)& Pol deg ($\%$) & Pol angle & MDP ($\%$)) \\
\hline\hline 
& \multicolumn{4}{c}{$\epsilon1$} &\\
\hline
$2.0$-$3.0$ &$0.51\pm0.7$ & $-0.32\pm0.7$ &  $<1.3$ & $-15.99^{\circ}\pm33.44^{\circ}$ & $2.11$  \\ 
$3.0$-$3.7$ &$-0.03\pm0.27$ & $0.19\pm0.27$ & $<0.38$ & $49.05^{\circ}\pm40.95^{\circ}$ & $0.83$  \\ 
$3.7$-$4.2$ &$0.25\pm0.24$ & $0.49\pm0.24$ & $<0.79$ & $31.3^{\circ}\pm12.63^{\circ}$ & $0.74$  \\ 
$4.2$-$8.0$ &$-0.7\pm0.31$ & $0.38\pm0.31$ & $<1.11$ & $75.74^{\circ}\pm11.08^{\circ}$ & $0.94$  \\ 
\hline 
& \multicolumn{4}{c}{$\epsilon2$} &\\
\hline
$2.0$-$3.0$ &$0.34\pm0.42$ & $-0.54\pm0.42$ & $<1.05$ & $-28.89^{\circ}\pm18.97^{\circ}$ & $1.27$  \\ 
$3.0$-$3.7$ &$0.58\pm0.16$ & $-0.04\pm0.16$ & $0.59\pm0.16$ & $-2.19^{\circ}\pm8.07^{\circ}$ & $0.5$  \\ 
$3.7$-$4.2$ &$0.6\pm0.15$ & $0.02\pm0.15$ & $0.6\pm0.15$ & $1.12^{\circ}\pm7.0^{\circ}$ & $0.44$  \\ 
$4.2$-$8.0$ &$-0.08\pm0.19$ & $-0.12\pm0.19$ & $<0.33$ & $-61.22^{\circ}\pm37.93^{\circ}$ & $0.56$  \\ 
\hline 
\hline
\end{tabular}
\end{table}

\begin{table}[h]
\caption{Polarization results from the W X-ray tube for $\epsilon1$. }
\label{table_W}
\centering   
\renewcommand{\arraystretch}{1.0}
\begin{tabular}{|c|c|c|c|c|c| } 
\hline\hline
Energy (keV) & Q/I ($\%$) & U/I ($\%$)& Pol deg ($\%$) & Pol angle & MDP ($\%$)) \\
\hline\hline 
& \multicolumn{4}{c}{$\epsilon1$} &\\
\hline
$2.0$-$3.25$ &$-0.86\pm0.5$ & $-0.31\pm0.5$ & $<1.42$ & $-80.01^{\circ}\pm15.65^{\circ}$ & $1.52$  \\ 
$3.25$-$4.25$ &$-0.51\pm0.23$ & $0.22\pm0.23$ & $<0.78$ & $78.17^{\circ}\pm11.95^{\circ}$ & $0.7$  \\ 
$4.25$-$5.25$ &$-0.43\pm0.21$ & $0.55\pm0.21$ & $0.7\pm0.21$ & $63.9^{\circ}\pm8.58^{\circ}$ & $0.64$  \\ 
$5.25$-$8.0$ &$0.27\pm0.19$ & $0.24\pm0.19$ & $<0.56$ & $20.89^{\circ}\pm14.92^{\circ}$ & $0.58$  \\ 
\hline 
\hline
\end{tabular}
\end{table}

\begin{table}[h]
\caption{Polarization results from the Rh X-ray tube for $\epsilon1$ and $\epsilon2$. }
\label{table_Rh}
\centering   
\renewcommand{\arraystretch}{1.0}
\begin{tabular}{|c|c|c|c|c|c| } 
\hline\hline
Energy (keV) & Q/I ($\%$) & U/I ($\%$)& Pol deg ($\%$) & Pol angle & MDP ($\%$)) \\
\hline\hline 
& \multicolumn{4}{c}{$\epsilon1$} &\\
\hline
$2.0$-$2.6$ &$-2.3\pm0.79$ & $4.42\pm0.79$ & $4.98\pm0.79$ & $58.73^{\circ}\pm4.53^{\circ}$ & $2.39$  \\ 
$2.6$-$3.2$ &$-1.49\pm0.46$ & $5.12\pm0.46$ & $5.33\pm0.46$ & $53.13^{\circ}\pm2.47^{\circ}$ & $1.4$  \\ 
$3.2$-$4.0$ &$-5.4\pm0.42$ & $14.13\pm0.42$ & $15.12\pm0.42$ & $55.45^{\circ}\pm0.79^{\circ}$ & $1.27$  \\ 
$4.0$-$7.0$ &$-7.38\pm0.29$ & $20.93\pm0.29$ & $22.2\pm0.29$ & $54.71^{\circ}\pm0.37^{\circ}$ & $0.88$  \\ 
\hline 
& \multicolumn{4}{c}{$\epsilon2$} &\\
\hline
$2.0$-$2.6$ &$-0.85\pm1.66$ & $-0.53\pm1.66$ & $<2.66$ & $-73.99^{\circ}\pm47.41^{\circ}$ & $5.04$  \\ 
$2.6$-$3.2$ &$1.74\pm0.94$ & $-5.96\pm0.94$ & $6.21\pm0.94$ & $-36.86^{\circ}\pm4.34^{\circ}$ & $2.85$  \\ 
$3.2$-$4.0$ &$6.16\pm0.86$ & $-13.63\pm0.86$ & $14.95\pm0.86$ & $-32.83^{\circ}\pm1.64^{\circ}$ & $2.6$  \\ 
$4.0$-$7.0$ &$8.93\pm0.59$ & $-20.61\pm0.59$ & $22.46\pm0.59$ & $-33.29^{\circ}\pm0.75^{\circ}$ & $1.79$  \\ 
\hline 
\hline
\end{tabular}
\end{table}

\begin{table}[h]
\caption{Polarization results from the Fe X-ray tube for $\epsilon1$ and $\epsilon2$. }
\label{table_Fe}
\centering   
\renewcommand{\arraystretch}{1.0}
\begin{tabular}{|c|c|c|c|c|c| } 
\hline\hline
Energy (keV) & Q/I ($\%$) & U/I ($\%$)& Pol deg ($\%$) & Pol angle & MDP ($\%$)) \\
\hline\hline 
& \multicolumn{4}{c}{$\epsilon1$} &\\
\hline
$2.0$-$3.0$ &$-5.82\pm0.66$ & $13.28\pm0.66$ & $14.5\pm0.66$ & $56.82^{\circ}\pm1.3^{\circ}$ & $1.99$  \\ 
$3.0$-$4.0$ &$-5.53\pm0.3$ & $15.44\pm0.3$ & $16.4\pm0.3$ & $54.86^{\circ}\pm0.52^{\circ}$ & $0.9$  \\ 
$4.0$-$5.0$ &$-8.59\pm0.31$ & $23.06\pm0.31$ & $24.61\pm0.31$ & $55.22^{\circ}\pm0.36^{\circ}$ & $0.93$  \\ 
$5.0$-$8.0$ &$-11.93\pm0.45$ & $32.5\pm0.45$ & $34.62\pm0.45$ & $55.08^{\circ}\pm0.38^{\circ}$ & $1.38$  \\ 
\hline 
& \multicolumn{4}{c}{$\epsilon2$} &\\
\hline
$2.0$-$3.0$ &$5.93\pm1.31$ & $-15.8\pm1.31$ & $16.87\pm1.31$ & $-34.72^{\circ}\pm2.23^{\circ}$ & $3.98$  \\ 
$3.0$-$4.0$ &$5.09\pm0.6$ & $-14.29\pm0.6$ & $15.17\pm0.6$ & $-35.2^{\circ}\pm1.13^{\circ}$ & $1.82$  \\ 
$4.0$-$5.0$ &$8.81\pm0.62$ & $-24.09\pm0.62$ & $25.65\pm0.62$ & $-34.96^{\circ}\pm0.69^{\circ}$ & $1.88$  \\ 
$5.0$-$8.0$ &$13.31\pm0.92$ & $-32.83\pm0.91$ & $35.42\pm0.91$ & $-33.97^{\circ}\pm0.75^{\circ}$ & $2.8$  \\ 
\hline 
\hline
\end{tabular}
\end{table}

\begin{table}[h]
\caption{The parameters of the fit (Applied voltage-V, angle between incoming electrons and outgoing photons-$\theta$) to the measured polarization as a function of photon energy. The real values are also shown to highlight the discrepancy. }
\label{fit_theoretical_func_table}
\centering   
\renewcommand{\arraystretch}{1.0}
\begin{tabular}{|c|c|c|c|c| } 
\hline\hline
X-ray tube & V$_{real}$ (kV) & $\theta_{real}$ & V$_{fit}$ (kV) & $\theta_{fit}$\\
\hline\hline 
Ca & 5.4 & 0$^\circ$  & $19.1\pm1.9$ & $-0.5^\circ\pm6.5^\circ$  \\
W  & 8.0 & 0$^\circ$  & $23.8\pm0.2$ & $0.05^\circ\pm0.01^\circ$ \\
Rh & 6.2 & 90$^\circ$ & $13.4\pm9.0$ & $28.9^\circ\pm23.6^\circ$  \\
Fe & 6.1 & 90$^\circ$ & $12.7\pm10.8$ & $35.3^\circ\pm40.7^\circ$ \\
\hline 
\hline
\end{tabular}
\end{table}

\clearpage

\clearpage


\vspace{1ex}


\end{spacing}
\end{document}